\documentclass[aps,twocolumn,showpacs]{revtex4}

\usepackage{graphicx}
\include{amssym}
\def\PSfig#1#2{\centerline{\scalebox{#1}{\includegraphics{#2}}}}

\begin{document}

\title{Analytic perturbation theory versus $1/N$ expansion 
       in the Gross-Neveu model} 

\author{A. A. Osipov}
\affiliation{Dzhelepov Laboratory of Nuclear Problems, JINR, 
         141980 Dubna, Russia}
\author{B. Hiller, A. H. Blin}
\affiliation{Centro de F\'{\i}sica Te\'{o}rica, Departamento de 
         F\'{\i}sica da Universidade de Coimbra, 3004-516 Coimbra, 
         Portugal}

\begin{abstract}
The $1/N$ expansion can be successfully used to calculate the Green
functions of the two-dimensional $O(2N)$ Gross -- Neveu model. In
parallel, the methods of analytic perturbation theory are also applied.
Comparing the results of these two calculations at leading order, we 
report on the surprising agreement between them. 
\end{abstract}


\maketitle


An analytic model for the QCD running coupling $\bar{\alpha}_s(Q^2)$
has been proposed in \cite{Shirkov:1997}. Since then, this concept 
evolved to become the method known as analytic perturbation theory
(APT) \cite{Shirkov:2007}. The model is essentially based on the old
idea of combining the renormalization group (RG) summation of the 
perturbative series of ultra-violet (UV) logarithms with analyticity 
in the $Q^2$ variable. This idea has been used previously in the
context of the QED ghost-pole problem \cite{Bogoliubov:1960}. Certain
reservations are in order when applying this idea to the QCD case.
Some of them have been pointed out in \cite{Shirkov:1997}. However,
this discussion did not address the pertinent aspect of QCD which 
relates to chiral symmetry breaking (ChSB) in the infra-red (IR)
region \cite{Colwit:1980}. It is commonly accepted that this 
nonperturbative phenomenon takes place in QCD. 
     
Recently attempts have been undertaken to apply a second order Bethe 
-- Salpeter formalism to the evaluation of the meson spectrum in QCD
\cite{Prosperi:2004}. The idea of broken chiral symmetry was taken
into account through an approximate solution of the appropriate Dyson
-- Schwinger equations. It has been shown that the meson spectrum is
well described if the analytic running coupling 
$\bar{\alpha}_{\mathrm APT}(Q^2)$ is used in the region 
$Q^2<1\,\mbox{GeV}^2$ , {\it i.e.} it was assumed there that except
for analyticity, the IR behaviour of this coupling does not require
any additional nonperturbative corrections, caused for instance by the 
dynamical ChSB effect in QCD.

The analyticity of the running coupling function is a general
requirement applied to any f\mbox{}ield theory, including the theories 
with dynamical ChSB. The latter however can possess an interesting 
property: there are models \cite{Reinhardt:1995} where the
analyticity of the running coupling is automatically fulf\mbox{}illed 
in the true ground state. On the other hand, the same coupling, being 
calculated in the false vacuum, possesses a Landau singularity in the
IR region of $Q^2$. At large momenta both calculations coincide. 

Let us imagine for a moment that one knows only the result of the
false vacuum calculation. It is natural in this case to use the APT
method. The corresponding analyticization procedure, based on the 
spectral representation of the K\" allen -- Lehmann type, gives rise
to a nonperturbative contribution, which ``cures'' the IR ghost-pole 
problem. This improvement in the low-energy part of the theory should 
nevertheless be consistent with the idea of spontaneously broken 
symmetry: improved calculations based on the
false vacuum should not deviate much from the result obtained from 
the true ground state. If deviations turn out to be big, one should 
assume that additional nonperturbative contributions must still be 
considered. On the contrary, if deviations are small the APT result
can be a good approximation even for theories with dynamical ChSB.

Certainly, it would be most tempting to use the APT result without 
introducing additional corrections due to ChSB, {\it i.e.,} supposing 
that the nonperturbative correction inherent in analyticity already 
emulates a major part of the ChSB ef\mbox{}fect, at least up to some 
low-energy scale. The attempt to extract the IR QCD coupling from the 
meson spectrum and to compare it with the APT result has been made 
recently in \cite{Prosperi:2006}.  

Here we show that there is a sensible way to discuss this point also
on pure theoretical grounds by applying  the APT method to a theory 
which possesses crucial properties of QCD and yet is solvable both in 
the UV and IR regions. To put it more precisely, following the 
soluble two-dimensional $O(2N)$-symmetric Gross -- Neveu (GN) model 
\cite{Gross:1974}, which is known to be a renormalizable and 
asymptotically free quantum f\mbox{}ield theory, we argue in this
letter that there is actually a remarkable agreement between the
running coupling obtained by the APT method and, independently, by the 
$1/N$ nonperturbative expansion. 

One should elucidate two points here. F\mbox{}irst, we compare only
the leading order results, believing that higher orders can be 
interesting if the f\mbox{}irst step happens to be successful. 
Second, to be precise one can speak only about accurate agreement in
the region of Euclidean momenta $Q^2 > \Lambda^2$ {\it i.e.,} above 
the perturbative Landau singularity. It would be too naive to expect 
that the simple analyticization procedure can perfectly describe all 
details of the nonperturbative dynamics. Nevertheless, our
calculations clearly show that this procedure is compatible with the 
fact that fermions acquire their masses dynamically. The same
conclusion has been made in \cite{Prosperi:2006} for QCD.
    
The ef\mbox{}fective coupling $\bar{g}^2(Q^2)$ of the four-fermion
interactions in the GN model, obtained after an RG resummation of the 
leading UV logs in the naive vacuum, {\it i.e.,} for massless
fermions, is described by the formula 
\begin{equation}
\label{RG}
   N\,\bar{g}_{RG}^{2\ (1)}(Q^2)=\frac{1}{\beta_0 \ln (Q^2/\Lambda^2
   )}\, . 
\end{equation} 
The l.h.s. is fixed as $N\to\infty$, this explains the presence of
the factor $N$. Here $\beta_0=1/2\pi$, $\Lambda$ is an RG invariant
scale, $\Lambda =\mu\exp [ - 1/ (2\beta_0g_\mu^2)],$ and $g_\mu$ is 
the coupling constant renormalized at a scale $\mu$ (a momentum 
cut-off scheme is used; dimensional regularization gives essentially 
the same result).

After imposing APT analyticity one obtains an expression similar to
that found for the QCD coupling $\bar{\alpha}_{\mathrm
  APT}^{(1)}(Q^2)$ (see Ref. \cite{Shirkov:1997})
at the one loop order in the region $Q^2>0$ 
\begin{equation}
\label{SF}
   N\,\bar{g}^{2\ (1)}_{\mathrm APT} (Q^2)=\frac{1}{\beta_0}\left(
   \frac{1}{\ln (Q^2/\Lambda^2)} + \frac{\Lambda^2}{\Lambda^2-Q^2}
   \right).
\end{equation}
The last term eliminates the unphysical Landau singularity at 
$Q^2=\Lambda^2$ from the perturbative RG improved result (\ref{RG}). 
This singularity results from the assumption that fermions are
massless in the vacuum. We do this on purpose. Our aim
at this stage is to obtain the invariant coupling $\bar{g}^2(Q^2)$ by 
the APT method without detailed description of the phase state of the
GN model. In some respect this resembles the QCD case where one obtains
the analogues of eqs. (\ref{RG}) and (\ref{SF}) neglecting details
related to the ChSB ef\mbox{}fect. As a possible alternative we
mention here a model for the QCD invariant coupling, which is based on
the hypothesis of finite gluon and quark masses \cite{Shirkov:1999}.

At present the main properties of the GN model are well studied. In 
particular, it is known that the $1/N$ expansion is a rather firm 
method to find that fermions are massive in the true vacuum 
\cite{Gross:1974}. Indeed, as has been stressed first by Witten 
\cite{Witten:1978}, the GN model possesses a low-temperature phase of 
the Berezinski -- Kosterlitz -- Thouless type \cite{Berezinsky:1970}. 
In this phase, the fermions have mass $M\neq 0$, just as if there were 
true spontaneous chiral symmetry breaking. It is fortunate (although 
perhaps just a coincidence) that the model exhibits in this way a 
feature that would be present in higher dimensions. Here we would like 
to stress only two important facts in favour of the $1/N$ expansion: 
(a) it yields the correct spectrum for the model \cite{Halpern:1975}; 
(b) the exact S-matrix obtained in \cite{Zamolod:1978} can be expanded 
in powers of $1/N$ correctly reproducing the known terms of the $1/N$ 
series (see also Ref. \cite{Berg:1978}). It shows that the $1/N$ 
expansion is a well convergent series. 

Thus, the $1/N$ expansion provides a solid basis for the calculation of
the fermion mass $M$, which results in $M=\Lambda$ at leading order 
(both the exact and $1/N$ results for the ratio $M/\Lambda$ can be
found in Ref. \cite{Forgacs:1991}). To obtain the RG invariant running 
coupling $\bar{g}^2(Q^2)$ one can use the mass dependent RG method
\cite{Bogoliubov:1955}. In the true ground state $(M\neq 0)$ one has
\cite{Gross:1974,Reinhardt:1995}
\begin{equation}
\label{GF}
   N\,\bar{g}^{2\ (1)}_{1/N} (Q^2)=\frac{1}{\beta_0}\left[ X\,
   \ln\frac{X+1}{X-1}\right]^{-1}, 
\end{equation}
where $X=\sqrt{1+4M^2/Q^2}$. Both expressions, namely, eqs. (\ref{SF}) 
and (\ref{GF}), have the same UV asymptotics. 

Note also that the fermion mass can be simply found by considering the 
one-loop ef\mbox{}fective potential $V(\phi_c)$ of the GN model as a 
function of a classical scalar field $\phi_c$ which represents the 
fermion-antifermion bound state 
\begin{equation}
   V(\phi_c)=\frac{N\phi_c^2}{4\pi}\left(\ln\frac{\phi_c^2}{\Lambda^2}
       - 1 \right).
\end{equation}
This function has two stationary points, $\phi_c=0$ and
$\phi_c=\Lambda$, and a positive second derivative, $V''(\Lambda
)=N/\pi$ (minimum), for the second case alone. These two solutions
correspond to the massless (false vacuum) and massive (true vacuum) 
fermion states.

\begin{figure}[t]
\PSfig{0.51}{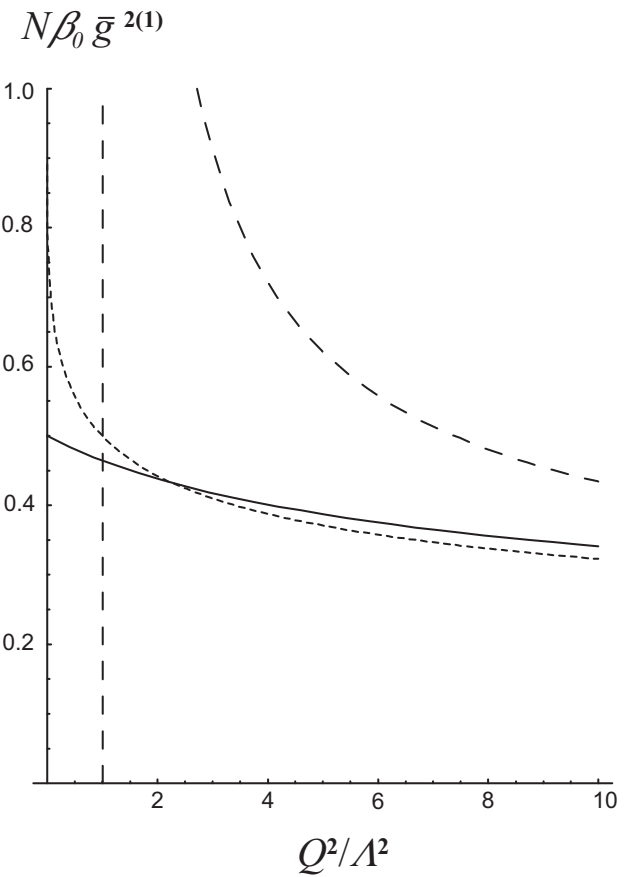}
\caption{\small The ef\mbox{}fective coupling of four-fermion
  interactions $N\beta_0\bar{g}^{2\,(1)}$ as a function of the ratio 
  $Q^2/\Lambda^2$. The curves represent the functions given by eq. 
  (\ref{SF}) (short-dashed curve), eq. (\ref{GF}) (solid curve) and
  the RG resummed perturbative result eq. (\ref{RG}) (long dashes). 
  Curves (\ref{SF}), (\ref{GF}) intercept the ordinate at $1$ and 
  $0.5$ respectively. The vertical dashed line corresponds to the 
  position of the Landau singularity at $Q^2/\Lambda^2=1$.} 
\label{fig1}
\end{figure}

The functions given by eqs. (\ref{RG}), (\ref{SF}), and (\ref{GF})  
are presented in f\mbox{}ig. \ref{fig1}. Our first observation is that 
the two functions $N\bar{g}^{2\ (1)}_{\mathrm APT} (Q^2)$ and 
$N\bar{g}^{2\ (1)}_{1/N} (Q^2)$ are quite close to each other in the 
interval $1< Q^2/\Lambda^2 < \infty$. In this domain the analytic 
correction agrees perfectly well with the well-established fact that
the fermions have a nonzero mass in the GN model. There is a sizeable 
dif\mbox{}ference in the results only at small values of $Q^2$, 
{\it i.e.,} in the IR region $0 < Q^2 < \Lambda^2$. Here, for
instance, the value $N\bar{g}^{2\ (1)}_{1/N}(0)=(2\beta_0)^{-1}$ is 
smaller than $N\bar{g}^{2\ (1)}_{\mathrm APT} (0)$ by a factor of two, 
and $\bar{g}^{2\ (1)}_{1/N}(0)=\bar{g}^{2\ (1)}_{\mathrm APT}(\Lambda^2)$.  
In this region, the results of the analytic model should be taken with
care. This is our second observation.

\begin{figure}[t]
\PSfig{0.37}{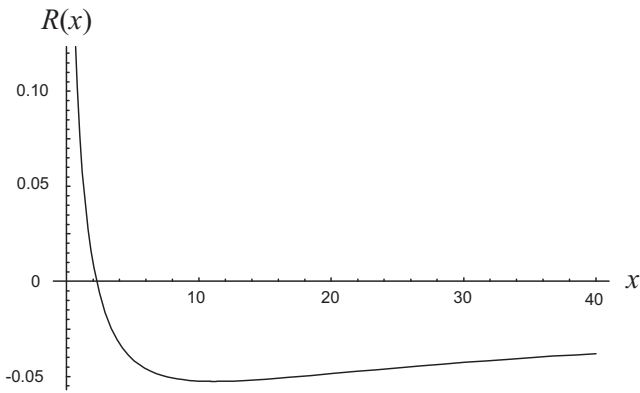}
\caption{\small The relative deviation of two quantities 
  $\bar{g}^{2\ (1)}_{\mathrm APT}(Q^2)$ and 
  $\bar{g}^{2\ (1)}_{1/N}(Q^2)$, described by eq. (\ref{ratio}),
  is shown as a function of the ratio $Q^2/\Lambda^2$.} 
\label{fig2}
\end{figure}

To compare the expressions obtained for the strength coupling, we
consider the ratio
\begin{equation}
\label{ratio}
   R(x)= \frac{\bar{g}^{2\ (1)}_{\mathrm APT}
          - \bar{g}^{2\ (1)}_{1/N}}{\bar{g}^{2\ (1)}_{1/N}}\, , 
             \quad x=\frac{Q^2}{\Lambda^2}\, ,
\end{equation}
which describes numerically the relative deviation of the APT result
from the $1/N$ prediction at point $x$. The function $R(x)$, shown in 
f\mbox{}ig. \ref{fig2}, decreases from $R(0)=1$ to $R(11.1)= -0.053$, 
where it reaches its minimum. The line $R=0$ is an asymptote for
$R(x)$ as $x\to\infty$. Inside the region $1<x<\infty$ the 
dif\mbox{}ference in values between the two approximations amounts to 
$|R|\simeq 5\%$ at most. 

The GN model with $U(1)$ chiral symmetry has a massless pseudoscalar 
state $\pi$ in the spectrum. The RG improved two-point Green function
of this state, $\Delta_\pi (Q^2)$, obtained at leading order in the 
$1/N$ expansion \cite{Gross:1974}, depends on the invariant coupling 
$\bar{g}^{2\ (1)}_{1/N}(Q^2)$ 
\begin{equation} 
\label{pion}  
   iN\Delta_\pi^{(1)} (Q^2)=\left(1+ \frac{4M^2}{Q^2}\right)\,
   N\bar{g}^{2\ (1)}_{1/N}(Q^2).
\end{equation}
The pseudoscalar propagator develops a pole at $Q^2=0$, as it should
be. On the other hand, its UV behaviour is described by formula
(\ref{RG}) in accordace with general requirements of chiral symmetry. 

\begin{figure}[t]
\PSfig{0.35}{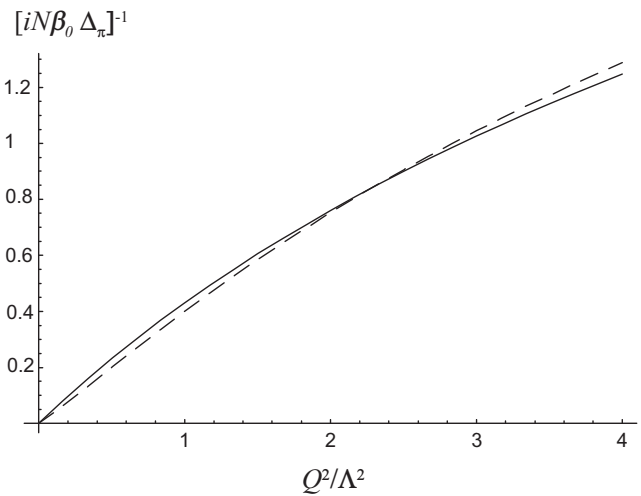}
\caption{\small The $1/N$ approximation for the inverse propagator
   of massless pseudoscalar (full curve) in comparison with the
   APT result (dashed curve) as functions of $Q^2/\Lambda^2$.}
 \label{fig3}
\end{figure}

Now, suppose that the strength $\bar{g}^{2\ (1)}_{1/N}(Q^2)$ in 
(\ref{pion}) is replaced by the APT coupling 
$\bar{g}^{2\ (1)}_{\mathrm APT} (Q^2)$. From our previous consideration  
one might expect to see the strongest ef\mbox{}fect from such 
replacement in the IR domain of $Q^2$. However, the physical IR 
singularity of eq. (\ref{pion}) at $Q^2=0$ dominates at small momenta, 
$0< Q^2 <\Lambda^2$, causing a substantial suppression of this 
ef\mbox{}fect. This renders the agreement between the two approaches 
compared in f\mbox{}ig. \ref{fig3} even better than between the 
quantities in f\mbox{}ig. \ref{fig1} (To avoid the singularity at 
$Q^2=0$ we plot the inverse function of (\ref{pion}) in f\mbox{}ig. 
\ref{fig3}). Indeed, the two functions given by eqs. (\ref{SF}) and 
(\ref{GF}), which may dif\mbox{}fer up to a factor of $2$ in the
region $0< Q^2 <\Lambda^2$, show almost the same behaviour after being 
multiplied by the term $(1+4\Lambda^2/Q^2)$. Therefore, although the 
low-energy behaviour of $\bar{g}^2$ given by the running coupling 
$\bar{g}^{2\ (1)}_{\mathrm APT} (Q^2)$ is less accurate as compared
with $\bar{g}^{2\ (1)}_{1/N} (Q^2)$, the analytic approach is still 
suitable for a reasonable description of the massless pseudoscalar
boson propagator in the IR region, provided the ChSB ef\mbox{}fects 
are correctly implanted. 

To conclude, we have compared here the leading order results of the
standard $1/N$ expansion of the GN model in the Euclidean region of 
momenta $Q^2>0$ with the corresponding results obtained by the methods of
analytic perturbation theory. These two alternative approaches
lead to dif\mbox{}ferent expressions for the ef\mbox{}fective RG 
invariant coupling constant $\bar{g}^{2\ (1)} (Q^2)$. In the 
f\mbox{}irst case the coupling $\bar{g}^{2\ (1)}_{1/N} (Q^2)$ is 
essentially the result which contains important information about the 
dynamics of ChSB at low energies. In the second case the analytic 
coupling $\bar{g}^{2\ (1)}_{\mathrm APT} (Q^2)$ is obtained without 
special regard to the ChSB ef\mbox{}fect, respecting only the basic 
principle of analyticity. The comparison shows that the nonperturbative 
contribution in $\bar{g}^{2\ (1)}_{\mathrm APT} (Q^2)$ corrects the RG 
improved perturbative result (at least up to the region of the Landau 
singularity at $Q^2=\Lambda^2$) in surprising accordance with the
general requirements of dynamical ChSB in the IR region. This is an 
interesting and a priori unexpected feature. We believe that our
result is generic for renormalizable theories with asymptotic
freedom and dynamical chiral symmetry breaking. Therefore, with due
care, it can also be applied to QCD, i.e. provided that the
ef\mbox{}fect of dynamical ChSB is indeed the essential mechanism 
governing the non-perturbative dynamics close to $Q^2=0$. If other 
details are important we expect new contributions beyound the GN
result to arise. 

\vspace{0.5cm}
{\bf Acknowledgements}
We are indebted to D. V. Shirkov for his interest, critical remarks,
and for clarifying discussions. This work has been supported in part
by grants provided by Funda\c c\~ao para a Ci\^encia e a Tecnologia, 
POCI/FP/63412/2005, POCI/FP/63930/2005 and POCI/FP/81926/2007. This 
research is part of the EU integrated infrastructure initiative Hadron 
Physics project under contract No.RII3-CT-2004-506078. 


\end{document}